\documentclass{ws-rv975x65_mod}
\usepackage{subfigure}     
\usepackage{ws-rv-thm}     
\usepackage{ws-rv-van}     
\usepackage{multirow}
\makeindex

\newcommand{\be}{\begin{equation}}
\newcommand{\ee}{\end{equation}}
\newcommand{\ba}{\begin{eqnarray}}
\newcommand{\ea}{\end{eqnarray}}
\newcommand{\bi}{\begin{itemize}}
\newcommand{\ei}{\end{itemize}}
\newcommand{\ben}{\begin{enumerate}}
\newcommand{\een}{\end{enumerate}}

\def \gsim{\mathrel{\vcenter
     {\hbox{$>$}\nointerlineskip\hbox{$\sim$}}}}

\def\gev{\mbox{$\mathrm{GeV}$}}
\def\tev{\mbox{$\mathrm{TeV}$}}
\def\pt{\mbox{$p_T$}}
\def\ptv{\mbox{$p_{T,V}$}}
\def\ptvsq{\mbox{$p^2_{T,V}$}}
\def\sintw{\ifmmode {\sin^2\theta_{\rm eff}^{\rm lept}} \else
  $\sin^2\theta_{\rm eff}^{\rm lept}$  \fi}
\def\sintc{\ifmmode {\sin^2\theta_C} \else $\sin^2\theta_C$ \fi}
\def\costc{\ifmmode {\cos^2\theta_C} \else $\cos^2\theta_C$ \fi}
\def    \=              {\;=\;}

\begin{document}
\chapter[$W/Z$ production at hadron colliders]{Production of
  electroweak bosons at hadron colliders: \\ theoretical
  aspects}

\author[]{Michelangelo L. Mangano}

\address{CERN, PH Department, TH unit \\
1211 Geneva 23, Switzerland \\
michelangelo.mangano@cern.ch
}

\begin{abstract}
Since the $W^\pm$ and $Z^0$ discovery, hadron colliders have provided
a fertile ground, in which continuously improving measurements and
theoretical predictions allow to precisely determine the gauge boson
properties, and to probe the dynamics of electroweak and strong
interactions. This article will review, from a theoretical
perspective, the role played by the study, at hadron colliders, of
electroweak boson production properties, from the better understanding
of the proton structure, to the discovery and studies of the top quark
and of the Higgs, to the searches for new phenomena beyond the
Standard Model.
\\
\\
{\it To appear in ``The Standard Theory up to the Higgs discovery - 60
years of CERN'', L. Maiani and G. Rolandi, eds. World Scientific.}
\end{abstract}
\vfill
\noindent
\\[1cm]
CERN-PH-TH-2015-286
\body

\section{Introduction}\label{mlm_sec1}
All bosons of the electroweak (EW) sector of the Standard Model (SM),
namely the gauge vector bosons $W^\pm$ and $Z^0$, and the scalar Higgs
boson $H^0$, have been discovered at hadron
colliders\cite{Arnison:1983rp,Banner:1983jy,Arnison:1983mk,Bagnaia:1983zx,Aad:2012tfa,Chatrchyan:2012ufa}.
This well known fact is sufficient to underscore in the strongest
terms the key role played by hadron colliders in the exploration of
the EW sector of the SM.

In hadron colliders, the physics of EW gauge bosons has many
facets. The abundant production rates via the Drell-Yan (DY)
process\cite{Drell:1970wh} enables significant measurements of their
properties, the best example being the so far unparalleled precision
of the determination of the $W$ boson mass, $M_W$, obtained at the
Tevatron\cite{Aaltonen:2013iut}.  The production of EW gauge
bosons in the decays of the top quark and of the Higgs boson,
furthermore, makes them indispensable tools in the study of the
properties of these particles. The presence of $W$ and $Z$ bosons in
the final state of a hadronic production process acts as a tag of the
underlying dynamics, singling out a limited number of production
channels, which can then be studied with great precision, due to the
experimental cleaness of the leptonic decay modes, and thanks to the
high accuracy achieved by the theoretical calculations. Last but not
least, $W$ and $Z$ bosons appear as final or intermediate states in
the decay of most particles predicted in theories beyond the SM
(BSM). Examples include the heavy bosons of new gauge interactions,
supersymmetric particles, or heavy resonances featured in models of
EW symmetry breaking alternative to the SM. The theoretical
study and the measurements of EW gauge bosons are therefore a
primary ingredient in the physics programme of hadron colliders.

It is impossible to provide, in this contribution, 
a complete historical overview of the
development of this field, and to properly acknowledge the main
contributions to both theoretical and experimental developments: on
one side there are too many to fit in these few pages; on the other,
the field is undergoing a continuous development, thanks to the
multitude of new data that are arising from the LHC and to the rapid
theoretical progress. Each of the topics briefly touched upon in this
review is examined in the theoretical and experimental literature with
a great degree of sophistication, and only an extended discussion
would fully address the complexity and ramifications of their
theoretical implications.  In this review, I shall therefore limit
myself to expose the great diversity of gauge boson physics in hadron
collisions, through the qualitative discussion of the main ideas and
results.  Furthermore, I shall only cover the physics of vector gauge
bosons, since the Higgs boson is covered in other chapters of this
book.  

I shall start from the general properties of inclusive $W$ and $Z$
production, focusing on the transverse and longitudinal dynamics and
on the implications for the knowledge of the partonic content of the
proton. I shall then discuss the phenomenological interest in the
production of multiple gauge bosons. Finally, I shall overview the
various mechanisms of associated production of gauge bosons and other
SM particles, namely jets and heavy quarks.

\section{QCD aspects of inclusive vector boson production}
The main feature of inclusive gauge boson production in hadronic
collisions is that the leading-order (LO) amplitude, describing the
elementary process $q\bar{q}^{(')}\to V$ ($V=W,Z$) is purely
EW.  The dynamics of strong interactions, at this order, only
enters indirectly through the parton distribution functions (PDFs),
which parameterize in a phenomenological way the quark and gluon
content of the proton.\footnote{For the overview of the principles and
  tools of perturbative QCD and of the parton models, which are
  relevant to the physics of hadronic collisions, I refer to the
  Chapter in this book by R.K. Ellis.}  At the large momentum scales
typical of gauge boson production ($Q\sim M_V$), higher-order
perturbative QCD corrections to the inclusive production are
proportional to $\alpha_s(Q)$ and are typically small, in the range of
10-20\%. They are known\cite{Hamberg:1990np,Harlander:2002wh} today to
next-to-next-to-leading order (NNLO), including the description of the
differential distributions of the boson and of its decay
leptons\cite{Anastasiou:2003ds,Melnikov:2006kv,Catani:2009sm,Gavin:2010az}, 
leaving
theoretical uncertainties from higher-order QCD effects at the percent
level. These results have been incorporated in full Monte Carlo
calculations including the shower evolution, to give a complete
description of the physical final
states\cite{Karlberg:2014qua,Hoeche:2014aia,Alioli:2015toa}.  
Next-to-leading-order (NLO) EW corrections are also
known\cite{Dittmaier:2001ay,Baur:2004ig,Baur:2001ze,Li:2012wna}, and play an
important role both for precision measurements, and in the production
rate of dilepton pairs at large \pt\ or with large mass, above the
TeV, where they can 
be larger than 10\%. Finally, progress towards a complete calculation
of the mixed ${\cal O}(\alpha_s\alpha)$ corrections has been recently
  reported in Ref.~\refcite{Dittmaier:2015rxo}.

When considering the first and second generation quarks that dominate
the production of $W$ and $Z$ bosons, their weak couplings, including
the CKM mixing parameters, are known experimentally with a precision
better than a percent. This exceeds the accuracy of possible
measurements in hadronic collisions, indicating that such measurements
could not be possibly affected, at this level of precision, by the
presence of new physics phenomena. They therefore provide an excellent
ground to probe to percent precision the effects of higher-order QCD
corrections and of PDFs\cite{Forte:2013wc}.
To be more explicit, consider the leading-order (LO) cross section given by:  
\be
d\sigma (h_1 h_2 \rightarrow V+X) = \int dx_1~ dx_2 \sum_{i,j} f_i(x_1,Q)~
f_j(x_2.Q)~ d\hat\sigma (ij\rightarrow V) \; ,
\label{eq:ppbar}
\ee
where $x_{1,2}$ are the fractions of the hadrons
momenta and $f_{i,j}$ are the corresponding distributions of (anti)quark
flavours $(i,j)$. In the case of $W$ production (a similar result
holds for the $Z$), the LO partonic cross
section is given by:
\be
\hat\sigma (q_i\bar q_j\rightarrow W) \= \pi~{\sqrt{2}\over 3}~\vert
V_{ij}\vert^2~G_F~M^2_W~\delta (\hat s - M^2_W) 
\= A_{ij} ~ M^2_W~\delta (\hat s - M^2_W)
\label{eq:DY}
\ee
Here $\hat s = x_1 x_2 S$ is the partonic centre-of-mass energy squared, and
$V_{ij}$ is the element of the Cabibbo--Kobayashi--Maskawa (CKM) matrix.

Written in terms of $\tau=x_1x_2$ and of the rapidity 
$y=\log [(E_W+p^{z}_W)/( E_W-p^{z}_W)]^{1/2} \equiv
\log (x_1 / x_2)^{1/2}$, the differential and total
cross sections are given by:
\ba
\frac{d\sigma_{W}}{dy} &=& \sum_{i,j} \;{\pi\, A_{ij}\over M^2_W}
\, \tau ~f_i(x_1)~f_j(x_2) \; , \quad
x_{1,2}=\sqrt{\tau}e^{\pm y}
\\
\sigma_{W} &=& \sum_{i,j} \;{\pi\, A_{ij}\over M^2_W}
\, \tau \int^1_\tau {dx\over x}~f_i(x)~f_j\left({\tau\over
x}\right)
\,\equiv \, \sum_{i,j} \;{\pi\, A_{ij}\over M^2_W}
\tau {L}_{ij} (\tau)
\ea
where the function ${L}_{ij} (\tau)$ is usually called {\em
  partonic luminosity}.
In the case of $u\bar d$ collisions, ${\pi\, A_{ij}\over M^2_W}\sim 6.5$nb.
It is interesting to study the partonic luminosity as a function of
the hadronic centre-of-mass energy. This can be done 
by taking a simple approximation for the parton densities. Using the
approximate behaviour $f_i(x) \sim {1/ x^{1+\delta}}$, with $\delta <
1$, one easily obtains:
\be
{L}(\tau) \= 
{1\over \tau^{1+\delta}}~ \log \left({1\over\tau}\right) \quad
\mathrm{and} \quad
\sigma_W \propto
\left({S\over M^2_W}\right)^\delta\,\log \left({S\over
M^2_W}\right)\; .
\ee
The gauge boson production cross section grows therefore at least
logarithmically with the hadronic centre-of-mass energy.

\begin{figure}[hb]
\centerline{
\includegraphics[width=0.5\textwidth]{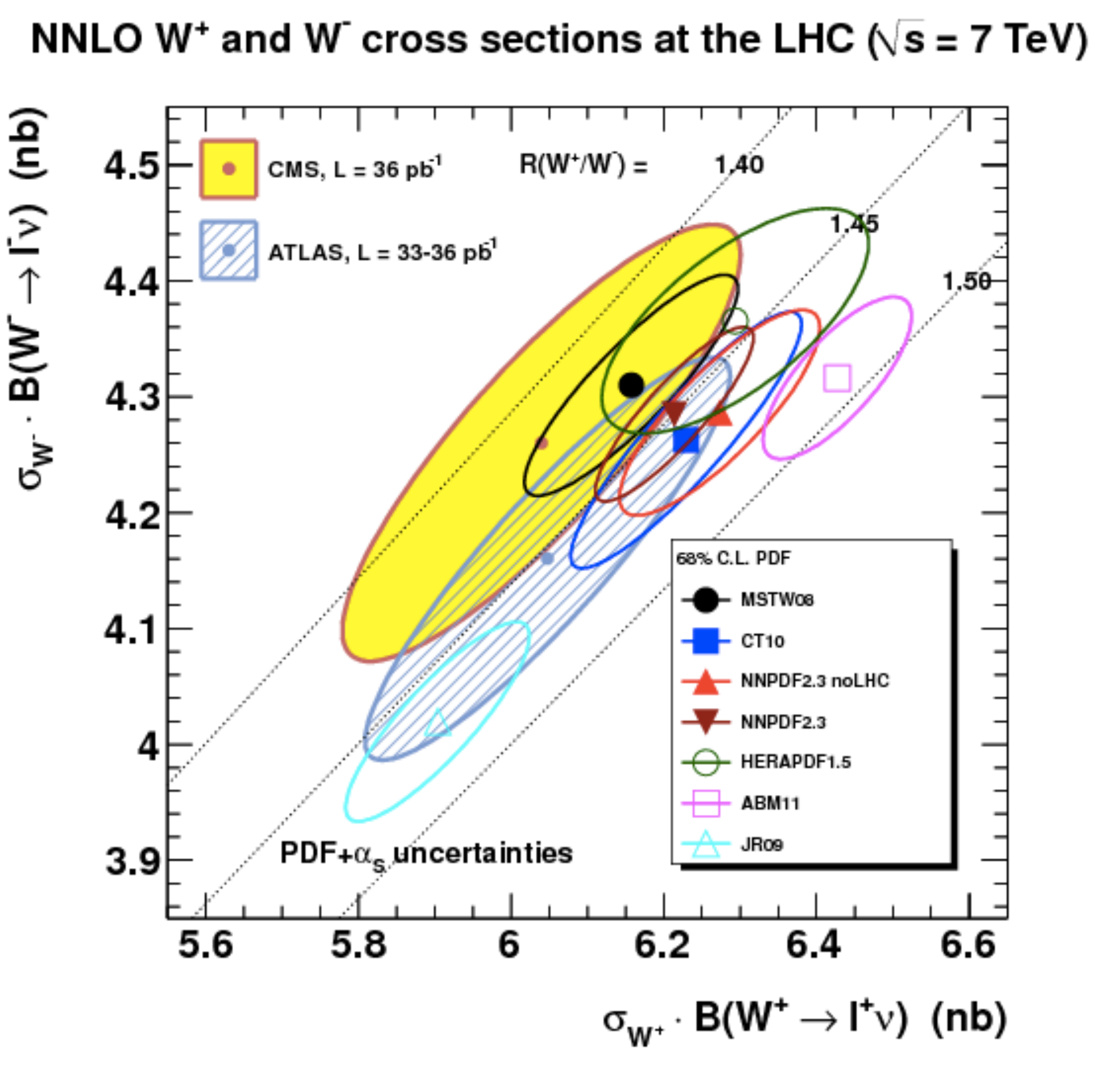}
\includegraphics[width=0.5\textwidth]{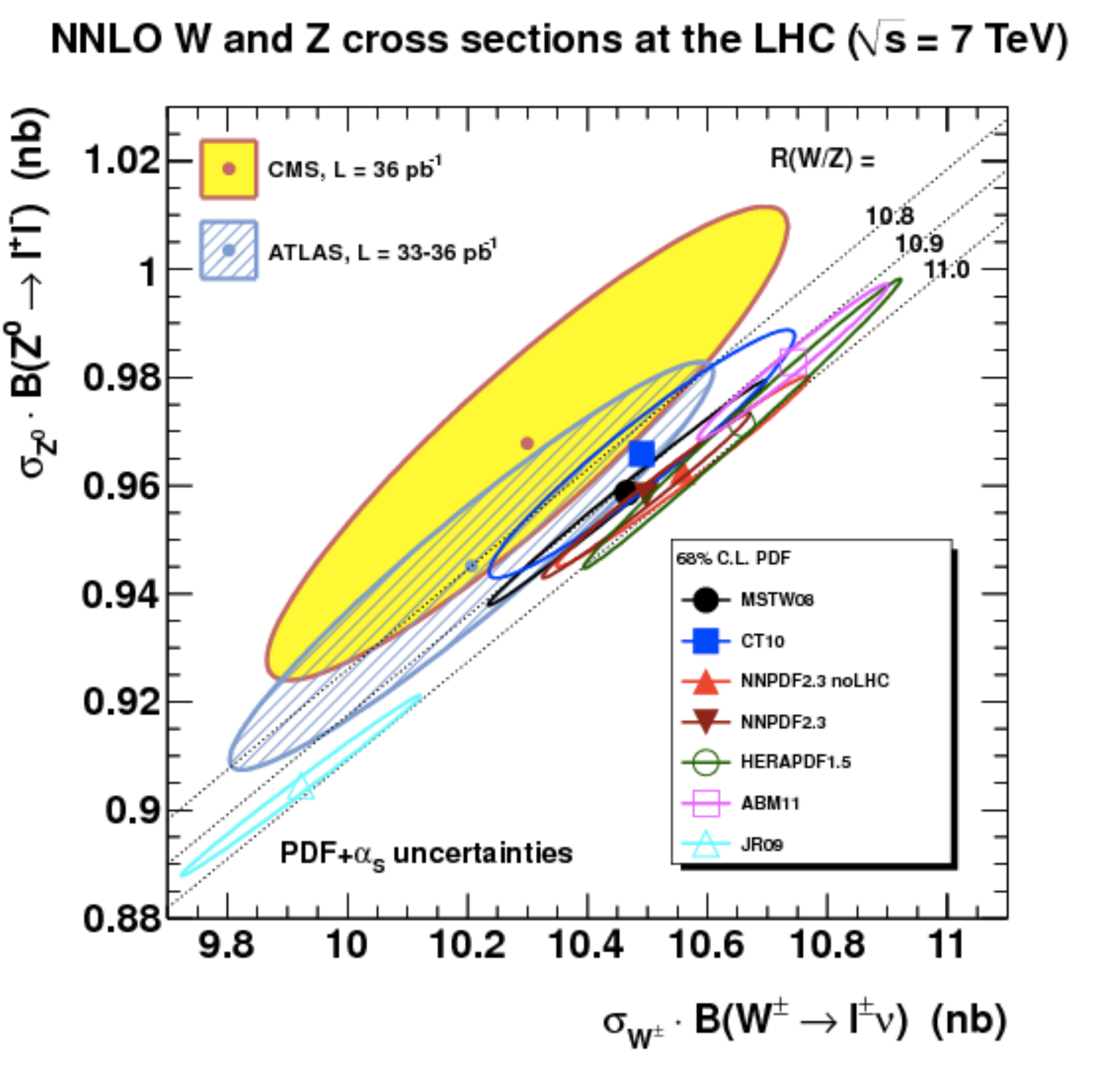}}
\caption{$W$ and $Z$ boson cross sections in $pp$ collisions at
  $\sqrt{S}=7$~TeV: ATLAS\cite{Aad:2011dm} and CMS\cite{CMS:2011aa}
  data, compared to NNLO predictions for various PDF
  sets\cite{Forte:2013wc}.}
\label{mlm_fig1}
\end{figure}

\subsection{Rapidity spectrum of $W$ and $Z$ bosons}
The features of the momentum distribution of vector bosons along the
beam direction ($z$) are mostly controlled by properties of the parton
PDFs. In particular, in the case of $W$ bosons, the differences
between the PDFs of up- and down-type quarks and antiquarks lead to
interesting production asymmetries. Since the measurement
of asymmetries is typically very accurate, due to the cancellation of
many experimental and theoretical uncertainties, 
these play a fundamental role in the
precision determination of quark and antiquark PDFs. 
Furthermore, the
production asymmetries are modulated by the parity violation of the
vector boson couplings, leading to further handles for the
discrimination of quark and antiquark densities, and inducing a
sensitivity to the weak mixing angle \sintw, which controls the vector
and axial components of $Z$ boson interactions.

\subsubsection{$W$ charge asymmetries}
\begin{figure}[hb]
\centerline{
\includegraphics[width=0.95\textwidth]{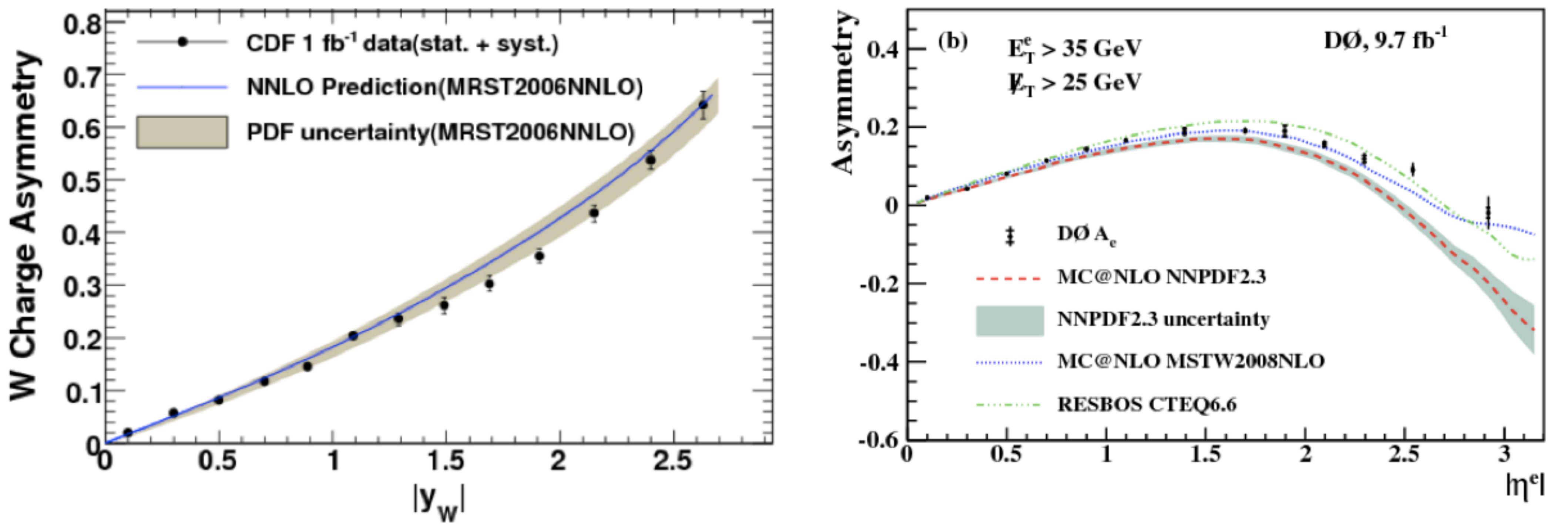}}
\caption{Production~\cite{Aaltonen:2009ta} (left) and
  leptonic~\cite{D0:2014kma} (right) charge asymmetries of $W$ bosons in
  $p\bar{p}$ collisions at the Tevatron, $\sqrt{S}=1.96$~TeV.}
\label{fig:Wasy_tev}
\end{figure}

For $p\bar{p}$ collisions, and assuming for simplicity
the dominance of $u$ and $d$
quarks, we have:
\ba
{d\sigma_{W^+}\over dy} &\propto& f^p_u(x_1) ~ f^{\bar p}_{\bar
d}(x_2) + f^p_{\bar d}(x_1) f^{\bar p}_u(x_2)  \\
{d\sigma_{W^-}\over dy} &\propto& f^p_{\bar u}(x_1) ~ f^{\bar p}_d
(x_2) + f^p_d(x_1) f^{\bar p}_{\bar u}(x_2) 
\ea
We can then construct the following charge asymmetry (using $f^{\bar
  p}_q = f^{p}_{\bar{q}}$ and assuming the
dominance of the quark densities over the antiquark ones, which is
valid in the kinematical region of interest for $W$ production at the 
Tevatron):
\be
A(y) \= -A(-y) \=
\displaystyle{{{d\sigma_{W^+}\over dy} - {d\sigma_{W^-}\over dy} \over
{d\sigma_{W^+}\over dy} + {d\sigma_{W^-}\over dy}}}
\sim
{f^p_u(x_1)~f^p_d(x_2) - f^p_d(x_1)~f^p_u(x_2)\over
f^p_u(x_1)~f^p_d(x_2) + f^p_d(x_1)~f^p_u(x_2)}
\ee
Setting $f_u(x) = f_d(x)R(x)$ we then get:
\be
A(y) \sim {R(x_1)-R(x_2)\over R(x_1)+R(x_2)}~,
\ee
which gives an explicit relation between asymmetry and the functional
dependence of the $u(x)/d(x)$ ratio. This ratio is close to 1 at small
$x$, where the quark distributions arise mostly from sea quarks, and
it increases at larger $x$, where the valence contribution
dominates. At positive $y$, where $x_1>x_2$, we therefore expect a
positive asymmetry. This is confirmed in the left plot of
Fig.~\ref{fig:Wasy_tev}, showing the asymmetry measured at the
Tevatron by the CDF experiment~\cite{Aaltonen:2009ta}, and compared to
the NNLO QCD
prediction\cite{Catani:2009sm,Catani:2010en,Gavin:2012sy,Li:2012wna} 
 and an estimate of the PDF uncertainty.
When measuring the charged lepton from $W$ decay,
the $W$ production asymmetry is however modulated by the $W$
decay asymmetry caused by parity violation. The squared amplitude for
the process $f_1 \bar{f}_2 \to W \to f_3 \bar{f}_4$ is proportional to
$(p_1 \cdot p_4)(p_2 \cdot p_3)$, where $f_{1,3}$ are fermions and
$f_{2,4}$ antifermions, of momenta $p_{1,\dots,4}$. In the rest frame
of this process, this is proportional to $(1+\cos\theta)^2$, where $\theta$
is the scattering angle between final- and initial-state fermions. 
The momentum of the final-state fermion, therefore, points 
preferentially in the direction of the  initial-state
fermion's momentum, $\cos\theta \to 1$. For $d\bar{u} \to W^- \to
\ell^-\bar{\nu}$ the charged lepton (a fermion) is more likely to move in the
direction of the $d$ quark, while for $u\bar{d} \to W^+ \to
\ell^+\nu$ the charged lepton (an antifermion) is more likely to move
  backward.  
The rapidity distribution of charged leptons is therefore subject to a
tension between the $W$ production asymmetry, which
at positive rapidity favours $W^+$ over $W^-$, and the decay asymmetry, which
at positive rapidity favours $\ell^-$ over $\ell^+$. The net result is
a distribution that changes sign, becoming negative at large lepton
rapidity. This is seen explicitly in the right plot of
Fig.~\ref{fig:Wasy_tev}, from the D0 experiment~\cite{D0:2014kma},
which also shows the great sensitivity of this quantity to different
PDF parameterizations, and the potential to improve their determination.

In $pp$ collisions,  assuming again the dominance of the first
generation of quarks and
$f^p_q(x)\gg f^p_{\bar{q}}(x)$ ($q=u,d$) at large $x$, the
$W$ charge asymmetry takes the form:\footnote{It goes without saying
  that in actual analyses the contributions of all quark and antiquark
  flavours are taken into account. At the LHC, in particular, 
  the contribution of strange and charm quarks is
  significant for the $W^\pm$ production rate, at the level of $\sim
  30 \%$.} 
\be
A(y)=A(-y) \sim {R(x_{max})-r(x_{min})\over R(x_{max})+r(x_{min})} \; ,
\ee
where $x_{max(min)}=max(min)(x_1,x_2)$ and $f^p_{\bar{u}}(x)=r(x)f^p_{\bar{d}}(x)$.
\begin{figure}[hb]
\centerline{
\includegraphics[width=\textwidth]{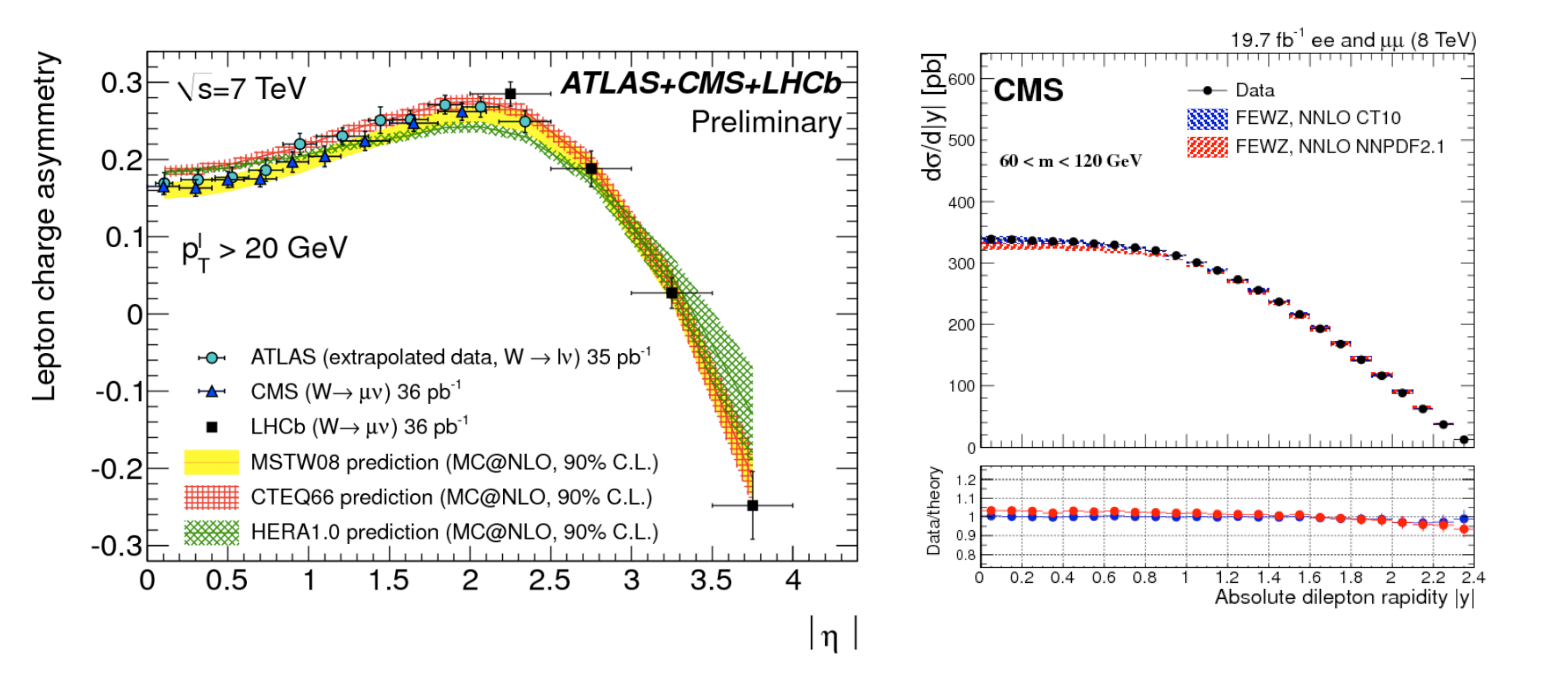}}
\caption{Left: leptonic charge asymmetries in $W$ production at the LHC
  ($\sqrt{S}=7$~TeV), extracted from the measurements of the
  ATLAS\cite{Aad:2011yna}, 
  CMS\cite{Chatrchyan:2013mza} and LHCb\cite{Aaij:2014wba}
  experiments. Right: $Z$ boson rapidity spectrum from
  CMS\cite{CMS:2014jea}, compared
  with NNLO predictions\cite{Li:2012wna}.}
\label{fig:Vrap}
\end{figure}
The extended rapidity coverage offered by the combination of the
ATLAS, CMS and LHCb detectors at the LHC, allows to fully exploit the
potential of asymmetry measurements as a probe of the proton
structure. This is highlighted in the left plot of 
Fig.~\ref{fig:Vrap}, which
summarizes the LHC experimental results for the lepton charge
asymmetry, obtained at $\sqrt{S}=7$~TeV,
compared to the theoretical predictions 
based on several sets of 
PDFs. In particular, notice the large spread of predictions in the
largest rapidity regions, spread to be reduced once these data
are included as new constraints in global PDF fits (see for example
Refs.~\refcite{Alekhin:2013nda,Ball:2014uwa,Harland-Lang:2014zoa,Dulat:2015mca}).
The PDF sensitivity can be further enhanced by considering the $W$
asymmetry at large rapidity in events produced in association with a
high-$p_T$ jet, as discussed in Ref.~\refcite{Farry:2015xha}.

\subsubsection{$Z$ rapidity spectrum and lepton charge asymmetries}
The measurement of the $Z$ rapidity spectrum is very accurate, due to
the precise reconstruction of its decay leptons. A comparison between
CMS data\cite{CMS:2014jea} and the NNLO theoretical calculation is
shown in the right plot of Fig.~\ref{fig:Vrap}. The agreement is
excellent, at the level of $\pm 5\%$, and on this scale one can detect
differences between the two choices of PDFs, CT10\cite{Gao:2013xoa} and
NNPDF2.1\cite{Ball:2011mu}, confirming the power of these measurements
to improve our knowledge of the quark distributions. Further inputs
will arise from the $Z$ production measurements performed
by LHCb\cite{Aaij:2015gna,Aaij:2015vua,Aaij:2015zlq}, in the range $2<y_Z<4.25$.

As discussed above, parity violation effects lead to particular
correlations in the decay directions of the final- and initial-state
fermions. For $Z^0$ production, these correlations depend
on the value of the weak mixing angle \sintw, which parameterizes the
relative strength of vector and axial couplings. In $e^+e^-$
collisions at the $Z$ pole, the measurement of these correlations is
in principle straightforward, since we know which of the initial state
particles is a fermion. The combination of such measurements, done at
LEP and SLC
using both leptonic and $b$-quark $Z$ decays,  and in particular using
at SLC polarized electron beams, led~\cite{ALEPH:2005ab} to the very
precise determination
of $\sintw=0.23153 \pm 0.00016$. These measurements remain
nevertheless puzzling, in view of a discrepancy between two of the most
precise inputs into the global average, namely LEP's measurement of
the forward-backward asymmetry of $b$ quarks ($A_{\rm FB}^{0,b}$),
$\sintw=0.23221 \pm 0.00029$, and SLD's measurement of the polarized
leptonic left-right asymmetry ($A_{\rm LR}$), $\sintw=0.23098 \pm
0.00026$. 

Due to the large statistics of $Z$ bosons, experiments at 
hadron colliders have the potential to contribute to these
measurements, and to address this puzzle. In practice, things are
complicated by the lack of information, on a event-by-event basis, on
which of the two intial-state partons is the quark, which is the
antiquark, and what is their flavour (note that the initial state could
also be $qg$, giving a different leptonic angular distribution). 
The problem is less severe in $p\bar{p}$ collisions than in
$pp$ collisions: in the former case, the most likely intial-state
configuration has the quark coming from the $p$ direction, and the
antiquark coming from the $\bar{p}$  direction. A residual ambiguity
remains, on whether the quark is of of $up$-type or $down$-type,
leading to a further slight dilution of the sensitivity. The
CDF and D0
experiments at the Tevatron have
presented the measurement of the weak mixing angle, from the analysis
of their full dataset of $Z$ decays (in the muonic channel for CDF,
and electronic channel for D0). CDF published\cite{Aaltonen:2014loa}
$\sintw= 0.2315 \pm 0.0010$.
D0 reports\cite{D0-sin2tw} $\sintw=0.21106 \pm 0.00053$, with
individual contributions of $\pm 0.0004$ from statistics and $\pm
0.0003$ from the PDFs. This is the most precise measurement from
hadronic colliders to date, with an overall uncertainty less than a
factor of 2 larger than the individual $A_{\rm FB}^{0,b}$ and 
$A_{\rm LR}$ determinations from LEP and SLD. These Tevatron
measurements are consistent with the overall LEP+SLD average. 

At the LHC, the extraction of the asymmetry is complicated by the
reduced discrimination between the $q\bar{q}$ and the $\bar{q}q$
initial states, since both beams are protons. This reduced sensitivity
is partly alleviated when considering events in which the $Z$ boson is
strongly boosted in either the forward or backward directions, since
in this case it is more likely that the quark moves in the direction
of the boosted $Z$, and that it is a $u$ rather than a $d$. The
extended rapdity coverage of the ATLAS and CMS experiments, and in
particular the very forward coverage of LHCb, allows these
measurements. The first result was reported by   
CMS\cite{Chatrchyan:2011ya}, based on the analysis of 1.1fb$^{-1}$
of data at 7~TeV: 
$\sintw=0.2287 \pm 0.0020({stat})\pm 0.0025({
  syst})$.
ATLAS\cite{Aad:2015uau} published a result based on the full
 4.8fb$^{-1}$ 7~TeV dataset: 
$\sintw=0.2308 \pm 0.0005({ stat})\pm 0.0006({
  syst})\pm 0.0009(\mathrm{PDF}) = 0.2308\pm 0.0012$. 
LHCb combined their measurements at both 7 and
8~TeV\cite{Aaij:2015lka}, to obtain
$\sintw=0.23142 \pm 0.00073({ stat}) \pm 0.00052({syst}) \pm
0.00056({theory)} = 0.2314 \pm 0.0011$, where the theoretical
uncertainty arises mostly from the PDF uncertainty. The size of the
PDF systematics in both ATLAS and
LHCb results underscores the importance of future progress that should
emerge from the ongoing PDF determination
programme\cite{Rojo:2015acz}, based on LHC
data. Current estimates suggest that the LHC experiments should
eventually reach systematics at the level of today's world average
uncertainty.

\subsection{Transverse momentum spectrum}
\begin{figure}[hb]
\centerline{
\includegraphics[width=0.6\textwidth]{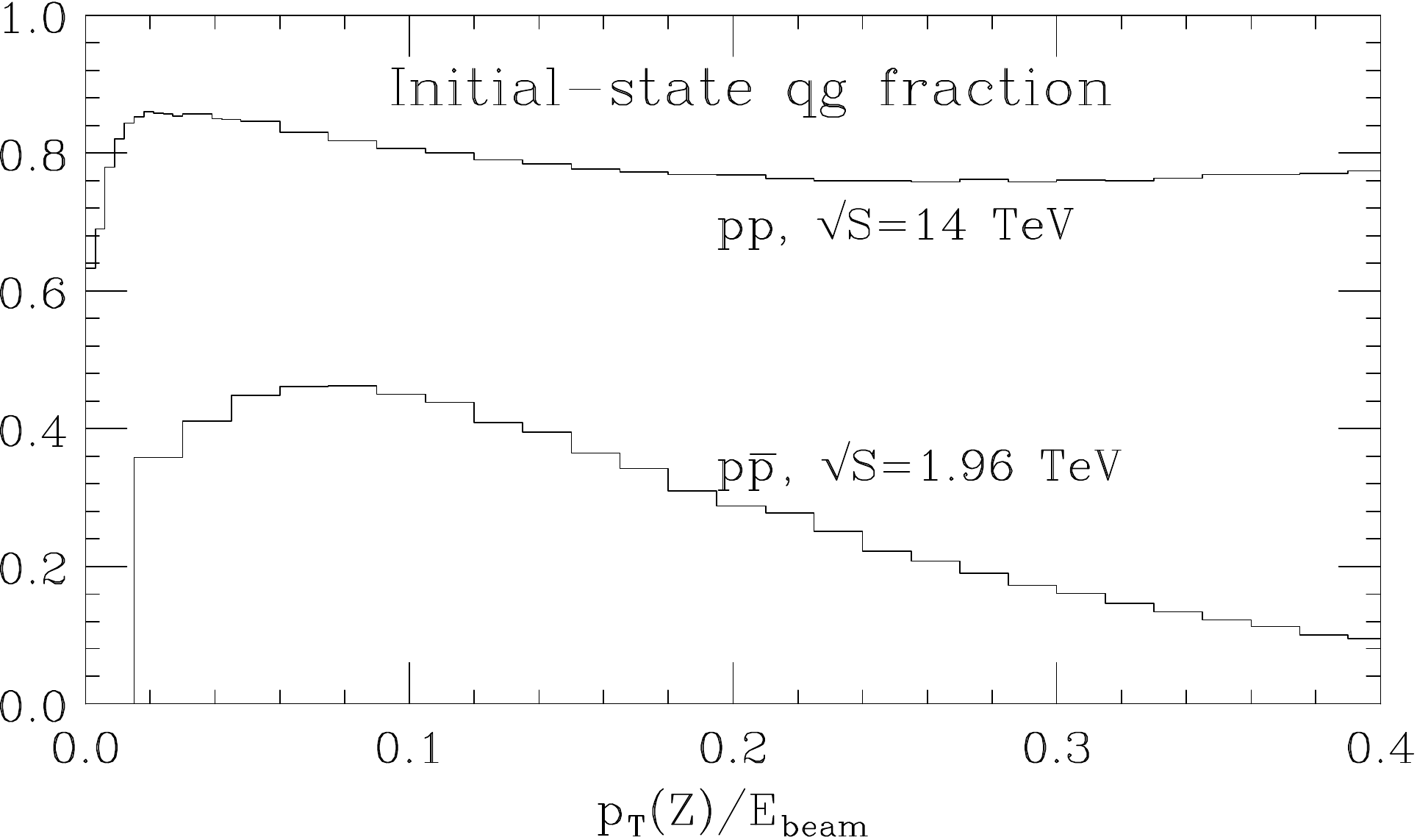}
}
\caption{Fraction of the $Z$ bosons of transverse momentum
  $p_T(Z)$ produced by the quark-gluon initial state, at the Tevatron and LHC.} 
\label{fig:ptz}
\end{figure}

When QCD corrections to inclusive gauge boson production are
considered, the most notable effect is the appearance of a transverse
momentum, \ptv. This is the result of parton-level processes such as
$q\bar{q} \to V g$ and $qg \to q V$. The former are typically dominant
in $p\bar{p}$ collisions, the latter in $pp$ collisions, as shown in
Fig.~\ref{fig:ptz}.  Depending on
the value of \ptv, relative to $M_V$, different dynamical and
theoretical issues are exposed, as summarized in this Section.

The spectrum at small \pt\ is dominated by the multiple emission of
soft gluons (soft with respect to the hard scale of the process,
namely $M_V$). This leads to corrections to $d\sigma/d\ptvsq$
proportional to $1/\ptvsq \alpha_S^n \log^m(M_V/\ptv)$ (where $n$ is
the number of soft gluons emitted, and $m\le 2n-1$), which need to be
resummed\cite{Dokshitzer:1978hw,Parisi:1979se,Curci:1979bg,Collins:1981uk}. The
leading-logarithmic 
soft-gluon resummation has been implemented in the context of
the exact fixed-order NLO
calculation\cite{Ellis:1981hk,Altarelli:1984pt,Gonsalves:1989ar,Arnold:1988dp}, 
and by now it has been extended
to next-to-next-to-leading logarithmic accuracy (NNLL). For the most
recent results, and a review of the existing literature on
resummation, see Ref.~\refcite{Catani:2015vma}.

At the lowest end, where $\pt\sim {\cal O}(\gev)$, the comparison of data
with LO theoretical calculations has historically required
the introduction of a modeling for the non-perturbative Fermi motion
inside the hadron\cite{Balazs:1997xd}. 
Most recently, the inclusion of exact higher-order
perturbative effects up to the next-to-leading order and the
resummation\cite{Bozzi:2010xn,Becher:2011xn}
of leading and sub-leading logarithms of $\pt/M_V$ greatly
reduced the need to introduce a phenomenological parameterization of
Fermi motion\cite{Abazov:2010mk}.
 
The production dynamics for \ptv\ of ${\cal O}(M_V)$ and beyond is
mostly controlled by purely perturbative physics, in addition of
course to the required knowledge of the partonic densities of the
proton. The comparison of data with theoretical predictions can
therefore be used to improve the determination of the PDFs.  The QCD
corrections are known up to
NLO\cite{Ellis:1981hk,Gonsalves:1989ar,Arnold:1988dp}, and work is in
progress towards a full NNLO result.  In $pp$ collisions the dominant
process for high-\pt\ vector boson production is the Compton-like
scattering $q g\to q' V$, as shown in Fig.~\ref{fig:ptz} for the
$Z$ boson.  This makes this process particularly sensitive to the
gluon PDF, over a very large range of $x$ values, as discussed in
detail in Ref.~\refcite{Malik:2013kba}. Such measurements should lead
in the future to a more accurate determination of the gluon PDF, an
essential step towards improving the precision of theoretical
predictions for the total production rate of Higgs bosons.

\begin{figure}[hb]
\centerline{
\includegraphics[width=0.8\textwidth]{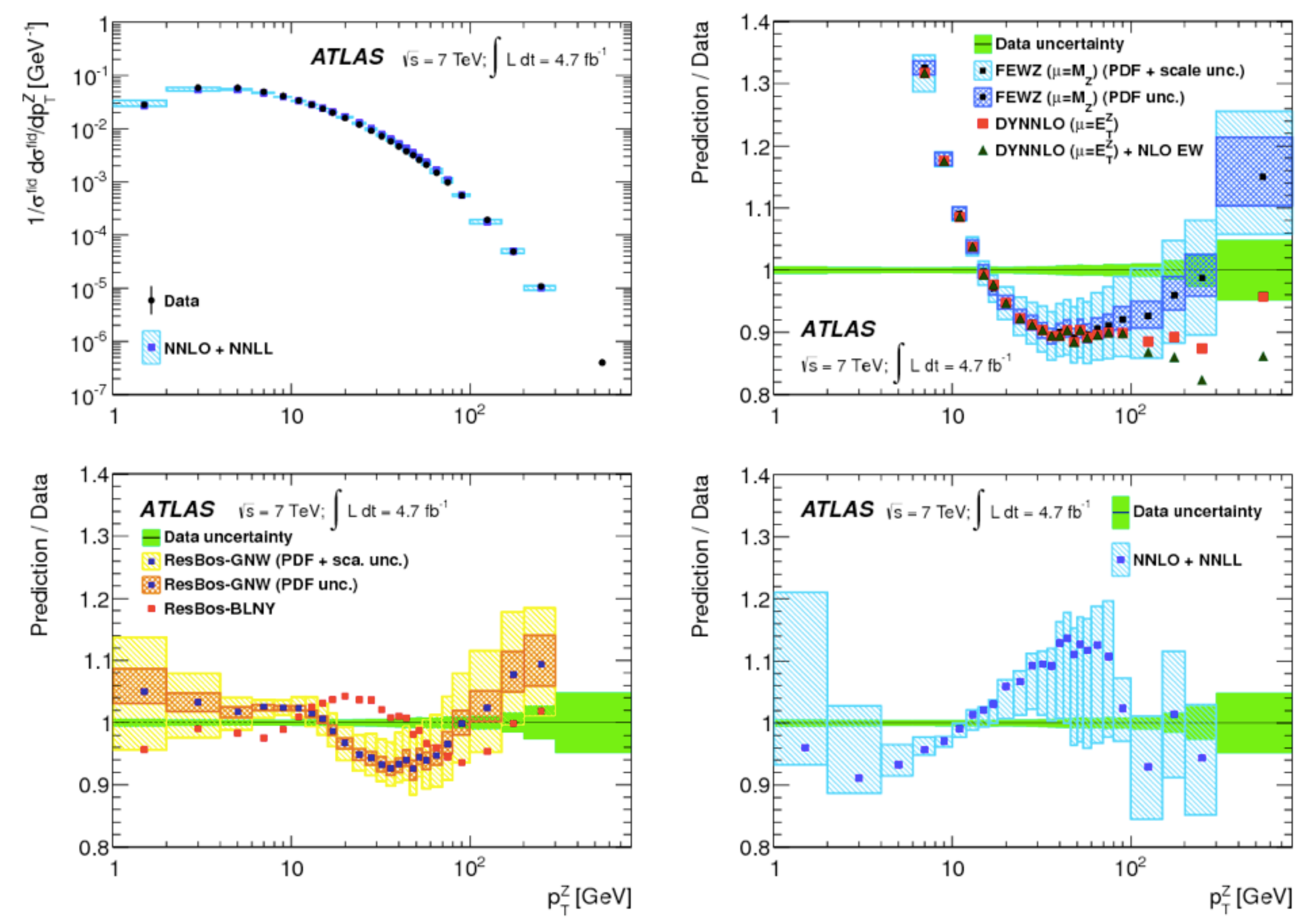}
}
\caption{$Z$ boson \pt\ spectrum measured by ATLAS\cite{Aad:2014xaa}
  (upper left), compared to various theoretical predictions:
  fixed-order NNLO (with different scales and with/without NLO EW
  corrections, upper right), resummed (N)NLO+(N)NLL 
(left~\cite{Balazs:1997xd,Guzzi:2013aja} and right\cite{Banfi:2012du}
lower plots).} 
\label{fig:ATLAS_Zpt}
\end{figure}
Figure~\ref{fig:ATLAS_Zpt} shows the recent ATLAS
results\cite{Aad:2014xaa}, compared to theory, for the $Z$ spectrum at
$\sqrt{S}=7~\tev$ (similar results at 8~TeV have been reported by
CMS\cite{Khachatryan:2015oaa}).  Notice the reach of the measurement,
extending up to \pt\ values of several hundred \gev, covering five
orders of magnitude in rate. The upper right plot compares, on a
linear scale, data and the results of NNLO
QCD\cite{Catani:2009sm,Li:2012wna} (NNLO here and in
fig.~\ref{fig:ATLAS_Zpt} refers to ${\cal
  O}(\alpha_s^2)$, namely NNLO for the inclusive rate, but NLO for
production at finite \pt).  In the region of $\pt\gsim
20~\gev$, where the effect of the small-\pt\ logarithms discussed
earlier is formally suppressed, data and theory agree to within 10\%.
At large \pt, 10\% differences arise when changing the functional form
of the renormalization scale $\mu$, from $\mu=M_W$ to
$\mu=\sqrt{M_W^2+\pt_W^2}$. Notice also the non-negligible effect of
NLO EW corrections\cite{Denner:2011vu}, which grow at large \pt.  
For smaller \pt\ values, where the fixed-order calculation is
insufficient, a better agreement with the data shape is obtained by
including the logarthmic resummation, an improvement included in the
theoretical predictions shown in the bottom two plots.
Overall, this comparison shows features in the pattern of the
comparison between data and theory, and between different theoretical
predictions, 
suggesting the need for
further improvements before these very precise data can be used, 
for example, for
improved determinations of the PDFs. 
Nevertheless, one should appreciate that the
overall scale of deviations, which are compatible with the quoted
uncertainties, is of order $\pm 10\%$, which remains quite impressive for a
process in hadronic collisions. 

\begin{figure}[hb]
\centerline{
\includegraphics[width=\textwidth]{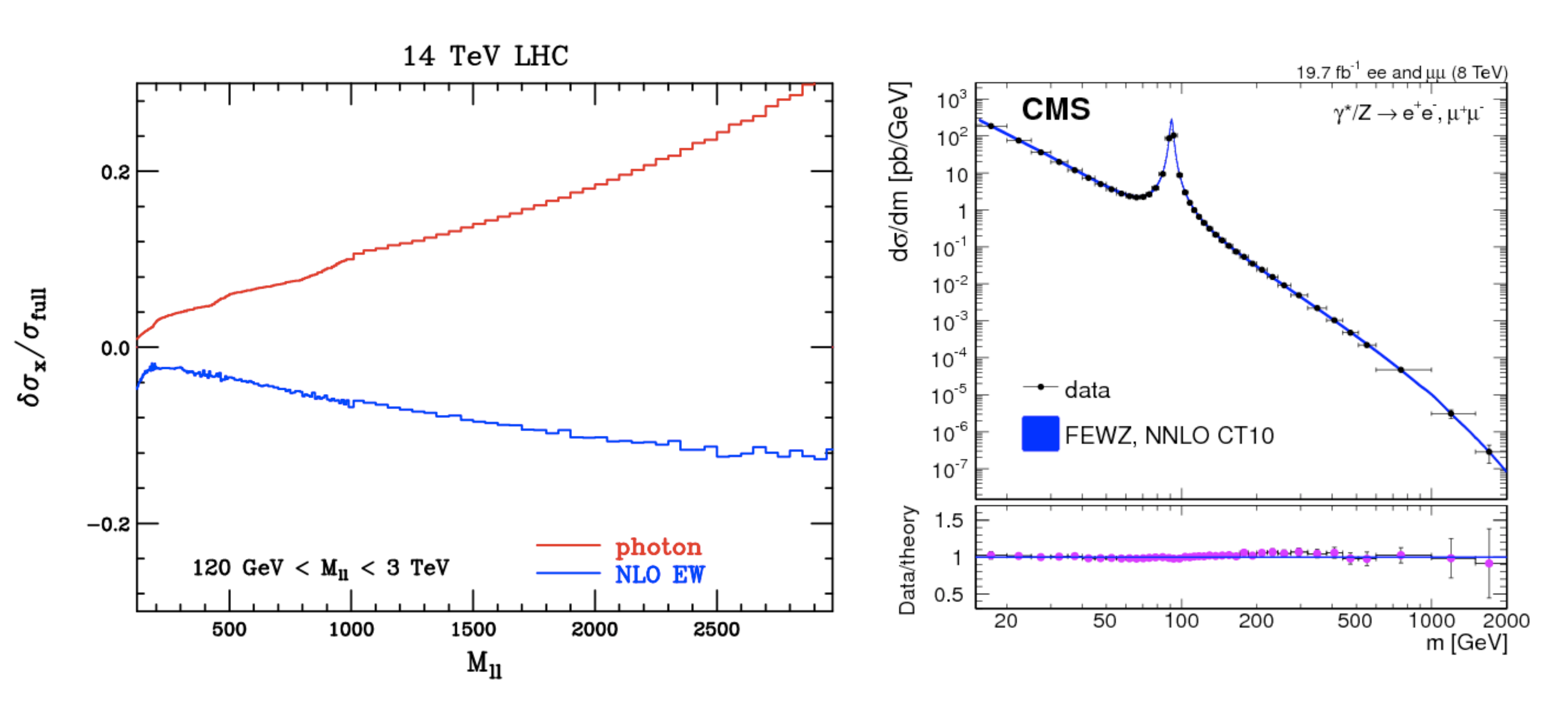}
}
\caption{Left: impact of EW and $\gamma\gamma\to \ell^+\ell^-$
  corrections\cite{Boughezal:2013cwa} on
  the DY mass spectrum at $\sqrt{S}=14$~TeV. Right: Drell-Yan mass spectra in $pp$ collisions at
  $\sqrt{S}=8$~TeV~\cite{CMS:2014jea}, compared against the 
  theoretical prediction\cite{Boughezal:2013cwa} including NNLO QCD
  and NLO EW corrections, with PDFs from Ref~\refcite{Gao:2013xoa}.} 
\label{fig:DYmass}
\end{figure}
\subsection{Off-shell gauge-boson production at large invariant mass}
The study of large-mass DY pairs is the primary probe in the search
for new interactions, characterized by the existence of heavy $Z'$ 
($W'$), and detectable as resonances (or jacobian peaks) 
in the $\ell^+\ell^-$ mass (or $\ell\nu$ transverse mass) spectra. 
This process, furthermore, tests the pointlike nature of quarks and
leptons or the possible existence of contact interactions, mediated
by heavy states beyond the reach of on-shell production. In this case,
BSM signals would appear as smooth deviations w.r.t. the SM behaviour
in the tails of the mass spectrum. Interpreting such deviations
requires a reliable control of the SM prediction, including the
precise knowledge of PDFs and of higher-order QCD and EW effects, including, as
recently emphasized in the literature, of purely QED processes
such as $\gamma\gamma \to \ell^+\ell^-$, which require as input the
knowledge of the photon density inside the proton. An example of the
impact of NLO EW and of the $\gamma\gamma$ corrections is
given in Fig.~\ref{fig:DYmass}, obtained in
Ref.~\refcite{Boughezal:2013cwa} using the photon PDF from the NNPDF
analysis\cite{Ball:2013hta}. 
Notice the large compensation
between the two opposite-sign contributions. Notice also that the
$\gamma\gamma$ channel is particularly large for this
observable, since the DY pair here is allowed to have small $p_T$
($\gamma\gamma$-induced final states are peaked at small $p_T$). The
contribution to other large-$Q^2$ DY configurations, such as inclusive
production at large $p_T$, is suppressed. 

The excellent agreement between theory and data, at the few percent
level, is shown in the right plot of 
Fig.~\ref{fig:DYmass}, from a recent analysis of the CMS
experiment~\cite{CMS:2014jea}. 

\section{Multiple production of vector bosons}
Pair production of gauge bosons includes contributions from channels
like $f\bar{f} \to \gamma/Z^* \to W^+W^-$ (with e.g. $f=e,q$), which
probe directly gauge boson self-interactions, and are sensitive to
deviations from the SM value of the relevant couplings (deviations
known as ``anomalous couplings''). Until recently the most accurate
studies of these couplings came from LEP2 data above the $WW$
threshold. In hadron colliders, one can explore a much broader range
of momentum scales and of off-shell configurations (e.g. probing
channels such as $q\bar{q}' \to W^* \to W \gamma$). The limited
statistics and kinematical reach available at the Tevatron is nowadays
largely surpassed at the LHC, whose sensitivity to anomalous couplings
is quickly overtaking that of LEP2. The LHC will also have sensitivity
to quartic gauge interactions, via the contributions to the $VV\to VV$
amplitude in the vector boson scattering processes $q q \to q q VV$.

Final states with gauge boson pairs are also fundamental signatures of
Higgs decays, $H\to ZZ^*$ and $H\to WW^*$.  The direct $VV$ production
is an important background to the isolation of these Higgs signals, as
well as to the signals of associated Higgs production, $pp\to VH$,
where Higgs decays such as $H\to b\bar{b}$ or $H\to \tau^+\tau^-$ are
subject to the background coming from the tail of the $Z^0$ invariant
mass distribution in $pp\to VZ^0$.

Furthermore, as a source of multilepton final states, multiple gauge
boson production is a potential background to a large number of BSM
searches, for example searches for the supersymmetric partners of
gauge and higgs bosons. 

Table~\ref{tab:multiV} collects the NLO production rates for all
processes with up to four massive gauge bosons in the final
state, taken from the comprehensive tabulation of NLO results for
high-multiplicity final states in Ref.~\refcite{Alwall:2014hca}, which
also lists the respective systematic uncertainties. 
\begin{table}[ht]
\tbl{Production cross sections (pb) in $pp$ collisions at 13~TeV for
  processes with multiple vector boson final states, from
  Ref.~\refcite{Alwall:2014hca}. }
{\begin{tabular}{@{}ccccccc@{}} \toprule
$W^\pm$         & $Z^0$      & $W^+W^-$   & $W^\pm Z^0$ & $Z^0Z^0$ 
& $W^+W^-W^\pm$ & $W^+W^-Z^0$ \\
\colrule
$1.7 \cdot 10^5$  & $5.4 \cdot 10^4$  & $1.0 \cdot 10^2$  & $4.5 \cdot
10^1$  
& $1.4 \cdot 10^1$  & $2.1 \cdot 10^{-1}$  & $1.7 \cdot 10^{-1}$  \\
\toprule
$W^\pm Z^0Z^0$ & $Z^0Z^0Z^0$ & $W^+W^-W^+W^-$ & $W^+W^-W^\pm Z^0$ & $W^+W^- Z^0Z^0$ 
& $W^\pm Z^0Z^0Z^0$ & $Z^0Z^0Z^0Z^0$ \\
\colrule
 $5.6 \cdot 10^{-2}$ & $1.4 \cdot 10^{-2}$  & 
$1.0 \cdot 10^{-3}$  & $1.2 \cdot 10^{-3}$  & $7.1 \cdot 10^{-4}$  & $1.2 \cdot
10^{-4}$  
& $2.6 \cdot 10^{-5}$   \\
\botrule

\end{tabular}
}
\label{tab:multiV}
\end{table}
Most of these processes should be eventually measurable at the
LHC.  The ratio $\sigma(W^\pm)/\sigma(Z^0) \sim 3$ (see
Table~\ref{tab:multiVratios}) turns out to be rather independent of
the energy, and of whether we consider $pp$ or $p\bar{p}$
collisions. When considering leptonic final states, this 
leads to the well know factor of $\sigma(W^\pm)BR(W\to \ell
\nu)/\sigma(Z^0)BR(Z\to \ell^+\ell^-)\sim 10$, which was observed
at the S$\bar{p}pS$, at the Tevatron and at the LHC.  This ratio
reflects primarily the nature and value of the couplings of $W$ and
$Z$ bosons to the up and down quarks in the proton. When the number of
final-state gauge bosons increases, the relative emission rate 
of further $W$ or $Z$
bosons gets closer to 1, as gauge boson selfcouplings become dominant
with respect to the couplings to initial state quarks. This is seen,
for example, in the first row of Table~\ref{tab:multiVratios}.
Notice, finally, that the cost of emitting additional gauge bosons
decreases with multiplicity: in part this is due to the larger number
of sources to couple to, in part to the reduced relative increase in
required energy (producing two gauge boson takes at least twice the energy than
producing one, while producing four takes only an extra 30\% more than
producing 3). 
 
\begin{table}[ht]
\tbl{Cross section ratios in $pp$ collisions at 13~TeV for
  processes with multiple vector boson final states.}
{\begin{tabular}{@{}cccc@{}} \toprule
$W^\pm/Z^0$   & $W^+W^-/W^\pm Z^0$ & $W^+W^-W^\pm /W^+W^-Z^0 $ 
& $W^+W^-W^+W^- /W^+W^-W^\pm Z^0 $ \\
\colrule
3.1 & 2.2 & 1.2 & 0.8 \\ 
\toprule
$W^+W^- / W^\pm$ &
$W^\pm W^+W^- / W^+W^- $ & 
$W^+W^- W^+W^-/ W^+W^- W^\pm$ & 
 \\
\colrule
$0.6 \cdot 10^{-3}$ &
$2.1 \cdot 10^{-3}$ &
$4.8 \cdot 10^{-3}$ &
 \\
\botrule
\end{tabular}
}
\label{tab:multiVratios}
\end{table}

At this time, the only final states measured with large statistics are those
with two bosons (for a complete phenomenological study of boson pair
production at the LHC, see e.g. Ref.~\refcite{Campbell:2011bn}). 
In the case of $W^+W^-$ production, in particular,
the statistical uncertainties of the LHC measurements at 8 TeV are already
about half the size of the systematics ones. 
Table~\ref{tab:VV} summarizes the status of
comparisons between data and theory for this channel. The consistency
of theoretical predictions and data is greatly improved by the
inclusion of the NNLO results~\cite{Gehrmann:2014fva}.\footnote{A similar
pattern is observed in the case of the $V\gamma$ production cross
sections, where the very recent completion of the NNLO (${\cal O}(\alpha_s^2)$)
calculations\cite{Grazzini:2015nwa} 
has improved the agreement with LHC
data\cite{Aad:2013izg,Khachatryan:2015kea}.} 
While compatible with uncertainties, some small
discrepancy is nevertheless present at 8~TeV, even between
experimental results. New
measurements at 13~TeV, and in particular the potentially more accurate
measurement of cross section ratios~\cite{Mangano:2012mh} between 13
and 8~TeV, will certainly clarify the whole picture.

On the theoretical side, note that at ${\cal O}(\alpha_s^2)$ a new
contribution appears, namely $gg\to WW$, mediated by a quark loop
(since this $gg$ channel enters for the first time at this order, its
description is referred to as LO even though it enters through a loop
diagram).  Its size is significant, due to the large $gg$ luminosity,
and contributes toward improving the agreement between theory and
data. The NLO correction to this new channel  (therefore of ${\cal
  O}(\alpha_s^3)$) has recently been computed\cite{Caola:2015rqy}.  At
the LHC, the correction relative to the LO $gg\to WW$ 
process can be
large, up to 50\%, leading to a further increase of the total cross
section by about 2\%. But the size of the cross section depends
strongly on the kinematical cuts applied to the final
state\cite{Caola:2015rqy}, an element that should be taken into proper
account in the comparison with the experimental measurements.  This
underscores the complexity of such high-precision tests of QCD
dynamics, but it is encouraging that continuous progress is taking
place in improving the theoretical calculations.

\begin{table}[ht]
\tbl{$W^+W^-$ cross sections measured in $pp$ collisions at 7 and 8
  TeV. The first three measurements include the Higgs
  contribution, the fourth one subtracts it. 
 NLO and NNLO theoretical predictions are from Ref.~\refcite{Gehrmann:2014fva}. 
The $gg\to WW$ process is included only in the NNLO contribution, and
the Higgs contribution\cite{Heinemeyer:2013tqa}, to be added to the
NNLO result for the comparison with the data in the first three rows, 
is shown separately.}
{\begin{tabular}{@{}ccccc@{}} \toprule
Experiment  & Data (pb) & NLO & NNLO & $gg\to H\to WW^*$ \\
\colrule
ATLAS\cite{Aad:2012oea} (7~TeV, incl. $H$) & $54.4 \pm 6.0 $ &
\multirow{2}{*}{$ 45.2{+1.7\atop -1.3}$} & 
\multirow{2}{*}{$49.0{+1.0\atop -0.9}$} & 
\multirow{2}{*}{$ 3.3{+0.2\atop  0.3}$}  \\ 
CMS\cite{Chatrchyan:2013yaa} (7~TeV, incl. $H$)   & $52.4 \pm 5.1$ 
&  &   & \\ 
\colrule
ATLAS\cite{ATLAS-WW-8TeV} (8~TeV, incl. $H$) & $71.4\pm 1.2_{stat} {+5.6 \atop -5.0}_{tot}$ &
\multirow{2}{*}{$54.8{+2.0\atop -1.6} $} & 
\multirow{2}{*}{$59.8{+1.3\atop -1.1} $} & 
\multirow{2}{*}{$ 4.1{+0.3\atop -0.3} $}  \\ 
CMS\cite{CMS:2015uda} (8~TeV, no $H$)   & $60.1 \pm
0.9_{stat} \pm 3.1_{th} \pm 3.5_{exp+lum}$ 
&  &   & \\ 
\botrule
\end{tabular}
}
\label{tab:VV}
\end{table}

\section{Associated production of vector bosons with jets and heavy
  quarks} The associated production of gauge bosons and
jets\cite{Campbell:2011zz} is a natural manifestation of higher-order
QCD corrections to inclusive production. The measurement of such final
states has a long history, starting from the CERN S$p\bar{p}$S
collider
experiments\cite{Arnison:1984iw,Bagnaia:1984dy,Ansari:1988sv}, which
highlighted their role as backgrounds to new physics, and as a probe
of $\alpha_S$.  Later studies at the
Tevatron\cite{Aaltonen:2007ip,Abazov:2013gpa} have been crucial to
test quantitatively the theoretical modeling, and to establish
background rates for the discovery and study of the top quark and for
the search of the Higgs boson. On the theory side, the last few years
have seen remarkable progress, with the NLO
calculations\cite{KeithEllis:2009bu,Berger:2009ep,Berger:2010vm,Berger:2010zx,Ita:2011wn,Bern:2013gka}
of processes with up to 5 jets, the inclusion of NLO
EW\cite{Denner:2009gj} corrections, and most recently of
NNLO QCD corrections to $W+1$~jet\cite{Boughezal:2015dva} and 
$Z+1$~jet\cite{Ridder:2015dxa} production.
\begin{figure}[hb]
\centerline{
\includegraphics[width=0.8\textwidth]{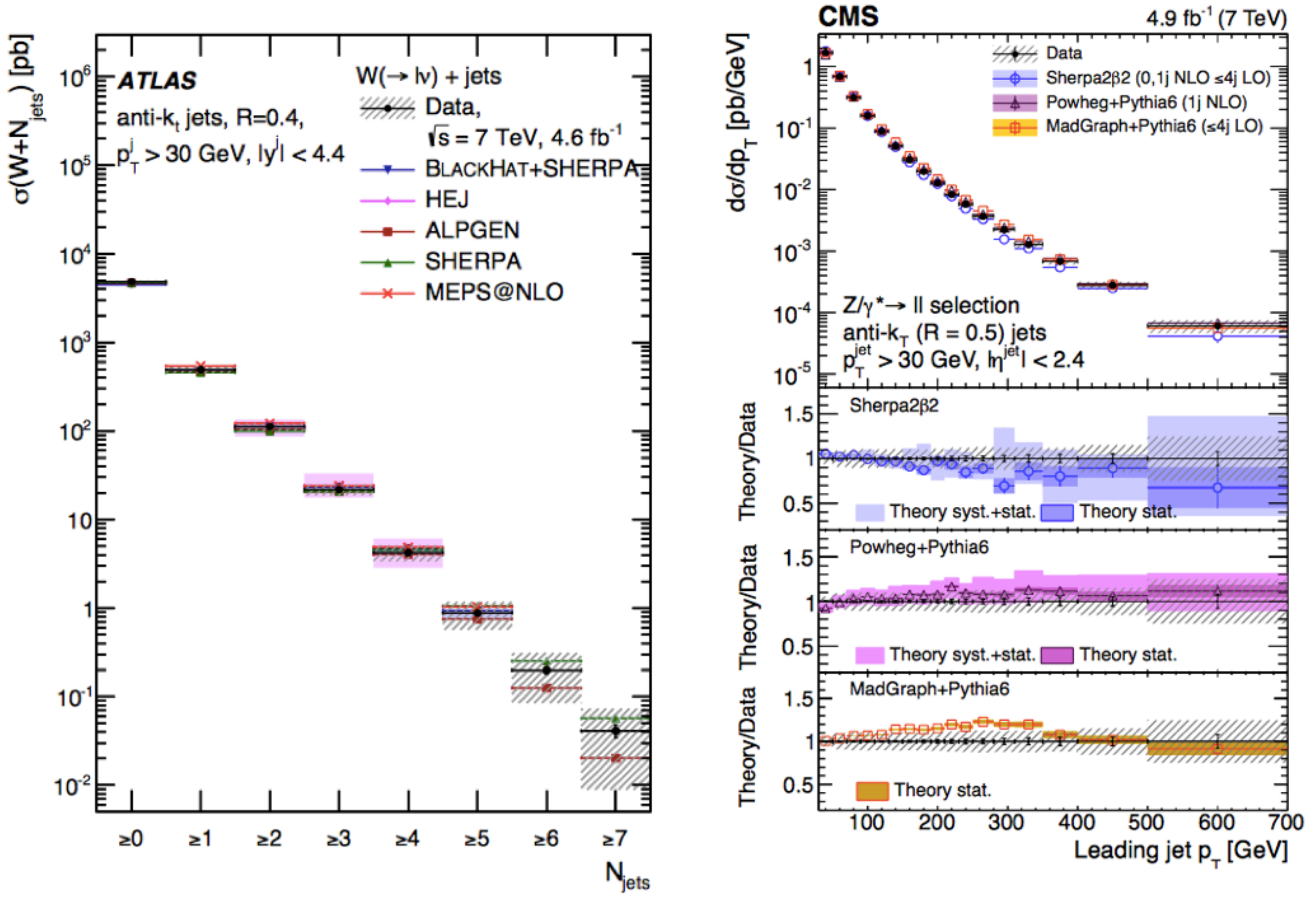}
}
\caption{Data vs. theory comparison for $V+$jets at the LHC. 
Left: $W+N$ jet rates at 7~TeV\cite{Aad:2014qxa}. Right:
  leading-jet \pt\ spectrum in $Z+$jets at 7~TeV\cite{Khachatryan:2014zya}.} 
\label{fig:Vjets}
\end{figure}

\begin{figure}[hb]
\centerline{
\includegraphics[width=0.6\textwidth]{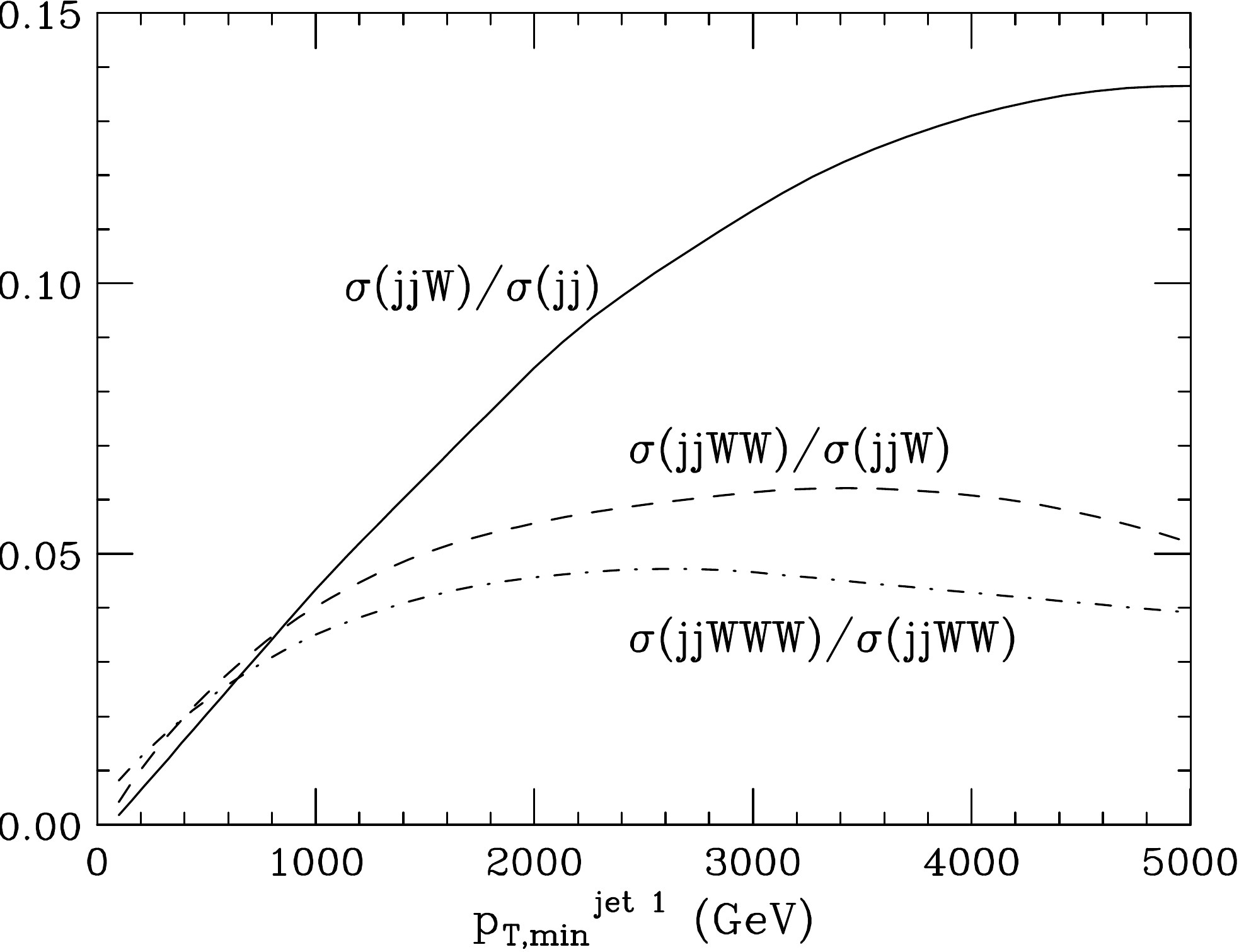}
}
\caption{Emission probability for $W$ bosons at 14~TeV, in events with 2 or more
  jets where the leading jet has $\pt>p_{T,min}$.} 
\label{fig:jjWfracs}
\end{figure}

The latest LHC
measurements\cite{Aad:2014qxa,Khachatryan:2014uva,Aad:2013ysa,Khachatryan:2014zya}
of $V+$ multijet production have reached multiplicities up to 7 jets,
with a precision and an agreement with theoretical calculations that,
at least for multiplicities up to 4 jets and for most kinematical
distributions, reach the level of $\pm (10-20)\%$. This is shown for
example in Fig.~\ref{fig:Vjets}.  Notice that a new feature of the production of
gauge bosons with jets emerges at the LHC, given the large jet
energies that can be reached: the probability of weak boson emission
increases, from the $10^{-3}$ level of inclusive QCD processes, up to
over 10\% for jet transverse momenta of several TeV (see
Fig.~\ref{fig:jjWfracs}). 

\subsection{$W+$charm quarks} 
At the LO, the $W+$charm 
cross section is proportional to $\sintc d(x)+\costc s(x)$, where
$d(x)$ and $s(x)$ are the PDFs of the down and strange quarks and
$\theta_C$ is the Cabibbo mixing angle. The Cabibbo-allowed process is
dominant, and allows the measurement of the strange quark
distribution\cite{Baur:1993zd,Giele:1995kr}. 
The difference in the production rate of $W^- c$ and
$W^+ \bar{c}$, after accounting for the small contribution of $d$ and
$\bar{d}$ quarks, is furthermore sensitive to the difference between $s(x)$
and $\bar{s}(x)$.\footnote{The assumption $s(x)=\bar{s}(x)$, which has been
used in the past in global PDF fits due to the lack of direct
experimental discriminating observables, is not respected at the NNLO in
the $Q^2$ evolution of PDFs.}  The first measurements at the
LHC\cite{Aad:2014xca,Chatrchyan:2013uja,Aaij:2015cha} have 
already led to useful constraints on PDF fits, but are still
statistics limited, and there is still large room for improvements.

\subsection{$V+Q\bar{Q}$, with $Q=c,b$} 
The associated production of vector bosons and heavy quark pairs is an
interesting SM process, which has particular relevance as leading
background to studies of the top quark, of the Higgs boson, and to
many searches for physics beyond the SM. While the case of $c\bar{c}$
and $b\bar{b}$ production are similar from the theoretical point of
view, we shall focus here on the case
of $b$ pairs, which has a larger phenomenological relevance, and for
which more experimental data are available.

$W b\bar{b}+jets$ production
gives rise to final states similar to those arising from $t\bar{t}$
production, and its presence, particularly at the Tevatron, was one of
the main hurdles in the top quark discovery and in its precision
studies. This is less so at the LHC, where its rate relative to the
$t\bar{t}$ signal is much smaller than at the Tevatron. $Vb\bar{b}$ is
an irreducible background to the associated production of gauge and
Higgs bosons, in the leading Higgs decay channel $VH\to Vb\bar{b}$. As
such, its understanding is nowadays the subject of many studies.

From the theoretical perspectve, the $Wbb$ and $Zbb$ processes are
rather different. In the former case, the only LO production channel
is $q\bar{q}' \to W b\bar{b}$. In the latter case, both $q\bar{q}$ and
$gg$ initial states contribute to the $Zb\bar{b}$ production. For
$q\bar{q}\to Vb\bar{b}$, the $b\bar{b}$ pair is produced by the
splitting of a final state gluon in the $q\bar{q}\to Vg$ process. The
difference between $q\bar{q}$- and $gg$-initiated processes is
particularly relevant when one considers final states where only one
$b$ jet is tagged: in the $gg\to Zb(\bar{b})$ case, in fact, there is
a large contribution induced by processes in which one of the gluons
in $gg\to Zbb$ undergoes a collinear splitting to $b\bar{b}$, and the
$b$ quark undergoes a hard scattering with the other gluon ($gb\to
Zb$), leading to the high-\pt\ tagged $b$-jet.  The $\bar{b}$ is
preferentially emitted at small \pt, covering a wide rapidity range,
and the integral over its full emission phase space leads to a large
logarithm. One could describe this process by associating this large
logarithm to the build up of a $b$ quark PDF inside the proton, and
the measurement of $Zb$ final states provides therefore a powerful
probe for the determination of the $b$ PDF. In the case of
$q\bar{q}\to Wbb$, on the other hand, the measurement of single
$b$-jet production receives comparable contributions from the cases
where the $\bar{b}$ is too soft to recostruct a jet, and cases in
which the pair produced by gluon splitting is collinear, and is merged
within the same jet. In the latter case, one exposes potentially large
logarithms $\log(p_T^{jet}/m_b)$.

For what concerns the comparison of theory and data from the
Tevatron\cite{Aaltonen:2008mt,Abazov:2013uza,Abazov:2015aha}
and from the
LHC\cite{Aad:2013vka,Chatrchyan:2013uza,Aaij:2014gta,Aad:2014dvb,Chatrchyan:2014dha,Aaij:2015cha}, 
the agreement of data for $V+b$-jet with NLO fixed-order perturbative
calculations\cite{Campbell:2005zv,Campbell:2006cu,Cordero:2009kv}
 is often marginal (possibly due to the presence of large
logarithms that call for improved resummed calculations), 
although consistent with the
uncertainties. A better agreement is typically found in the
comparisons with data where both $b$-jets are tagged. Future
measurements at the LHC, with larger statistics and better control on the
experimental systematics, will allow further improvements of the
theoretical modeling. 

\subsection{$V+t\bar{t}$} 
The associated production of $W$ and $Z$ bosons with a pair of top
quarks is a special case of the processes discussed in the previous
subsection. In many respects, the theoretical description of
$Vt\bar{t}$ production is however simpler: the mass of the top quark
and of the gauge bosons are both large and of comparable size, so that
we do not have the difficulties associated with the presence of
largely different scales. For example, there are no large logarithms to
be resummed, or assumptions to be made about the relevant heavy quark
density of the proton; furthermore, the prediction for the basic
process, namely the inclusive production of the heavy quark pair, is
much more precise for top quarks than for bottom or charm quarks. 
\begin{table}[ht]
\tbl{Production cross sections (pb) in $pp$ collisions at 13~TeV for various
  top quark and vector boson final states, from Ref.~\refcite{Alwall:2014hca}.}
{\begin{tabular}{@{}cccccc@{}} \toprule
$t\bar{t}$ & $t\bar{t} W^\pm$  & $t\bar{t} Z^0$  & $t\bar{t} W^+W^-$ 
& $t\bar{t} W^\pm Z^0$ & $t\bar{t} Z^0Z^0$  \\
\colrule
674  & 0.57 & 0.76  & $9.9 \cdot 10^{-3}$ 
& $3.5 \cdot 10^{-3}$ & $1.8 \cdot 10^{-3}$ \\
\botrule
\end{tabular}
}
\label{tab:ttV}
\end{table}

From the phenomenological perspective, the associated production with
top quarks has interesting features. To start with, at LHC energies 
and above these processes
are the main source of multiple gauge boson production. This is clear
from Table~\ref{tab:ttV}, which reports the total rates for several
final states involving top quarks and massive gauge bosons. 
Considering that each top and antitop quark produces a
$W$ boson in their decay, the comparison with the multi-$V$ rates given
in Table~\ref{tab:multiV} shows that final states with $W^+W^-$ pairs are more
likely to arise from $t\bar{t}$ production and decay than from direct EW
production. This is true as well of processes with production of
additional gauge bosons, e.g. $t\bar{t}Z$ versus $W^+W^-Z$. 
This fact should be taken into account when
extracting EW production rates from the data, and when estimating
multiboson backgrounds to BSM signals. 
\begin{figure}[hb]
\centerline{
\includegraphics[width=0.5\textwidth]{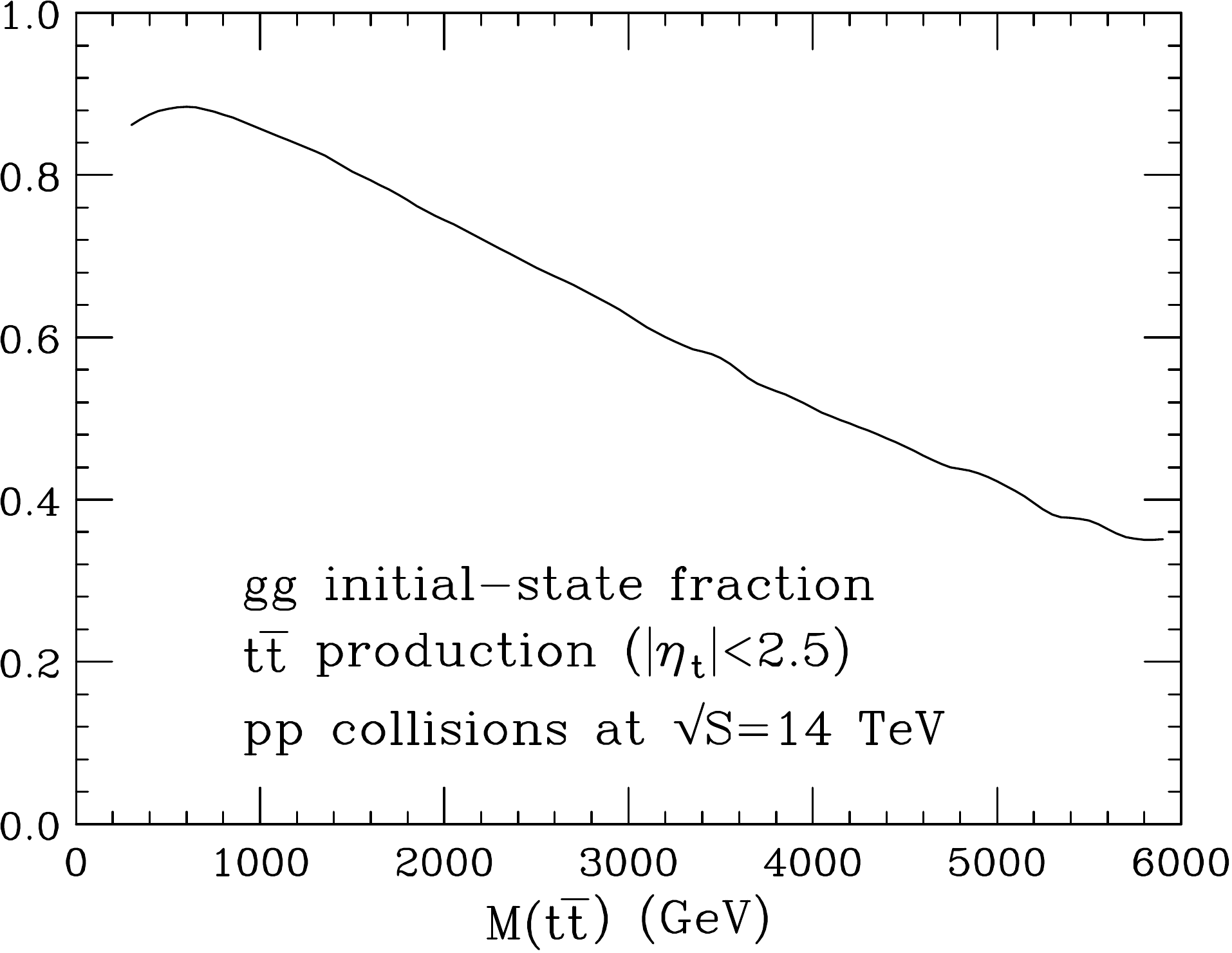}
\hfill 
\includegraphics[width=0.48\textwidth]{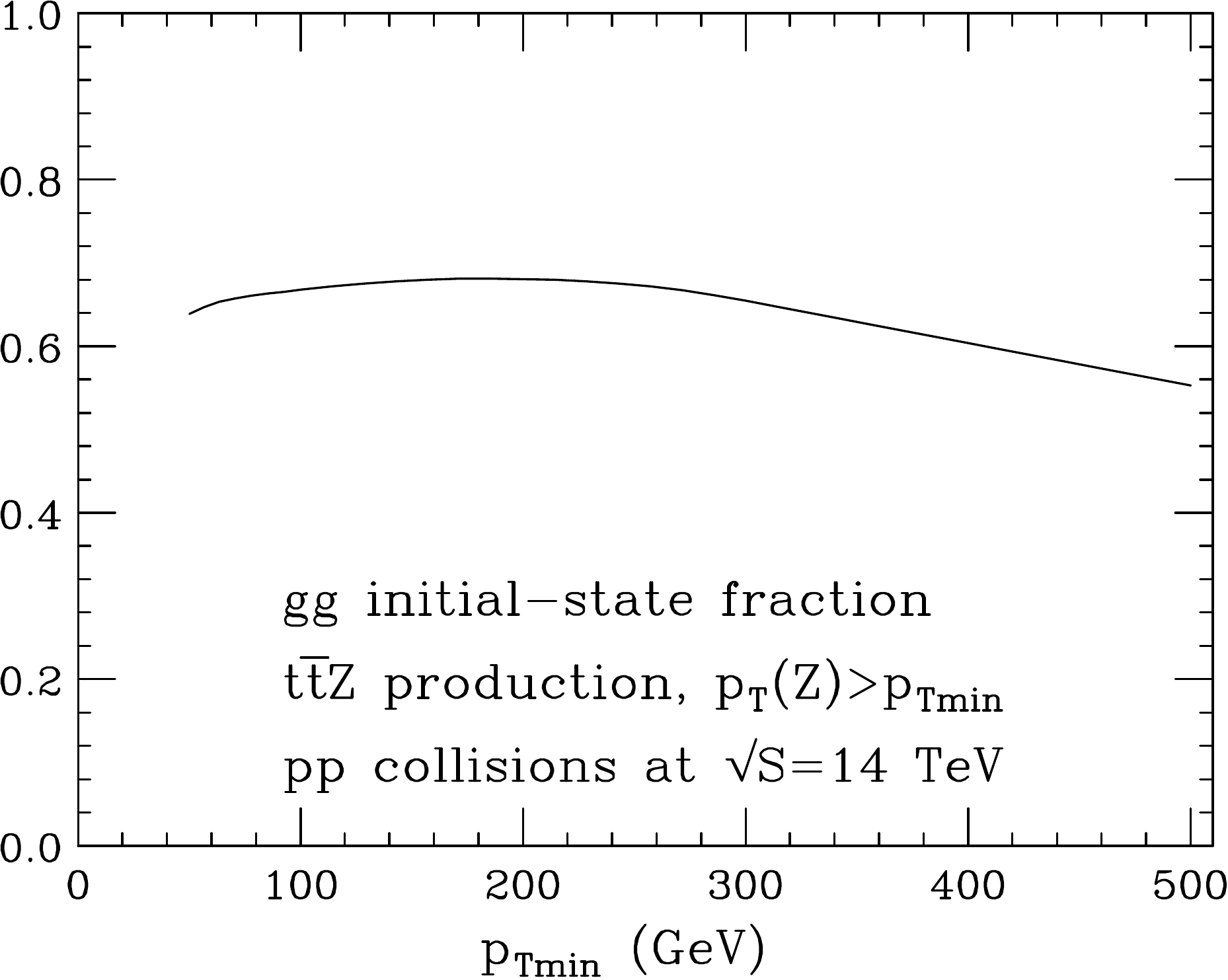}}
\caption{Initial state $gg$ fraction in inclusive production of
  $t\bar{t}$ (left) and $Zt\bar{t}$ (right), in $pp$ collisions at
  $\sqrt{S}=14$~TeV. } 
\label{fig:ttfrac}
\end{figure}

Another interesting observation is that, in $pp$ collisions at the LHC
energies and above, the production of a $Z$ boson is more frequent
than the production of a $W$ boson, contrary to the usual hierarchy of
$W$ vs. $Z$ production rates. The reason is that, at LO, the
$t\bar{t}W$ final state can only be produced starting from the
$q\bar{q}'$ initial state, while $t\bar{t}Z$ can be produced from both
$q\bar{q}$ and $gg$ inital states. The $gg$ fraction in $t\bar{t}Z$
production at 14~TeV is shown in Fig.~\ref{fig:ttfrac}, as a function
of $p_{T,Z}$. Since inclusive $t\bar{t}$ production is dominated by
the $gg$ channel (85\% of the rate at $\sqrt{S}=14$~TeV, see also the
left plot in Fig.~\ref{fig:ttfrac}), the emission of a $W$ is
suppressed with respect to the emission of a $Z$. This is shown
explicitly in Table~\ref{tab:ttV}, where the rates of various
processes with top quarks and gauge bosons are given. The usual
hierarchy in rate between $W$ and $Z$ production is restored for
associated production of $W^+W^-$ versus $ZZ$, when the $gg$ initial
state is active for both processes.

The above considerations have several corollaries. The production of a
$W$ boson, singling out the $q\bar{q}'$ initial state, allows to
scrutinize more closely the production mechanism $q\bar{q} \to
t\bar{t}$, which is otherwise suppressed at the LHC. This may be
useful~\cite{Maltoni:2014zpa} to enhance the sensitivity to possible
new physics at the origin of the forward-backward production asymmetry
reported at the
Tevatron~\cite{Aaltonen:2012it,Abazov:2014oea,Aguilar-Saavedra:2014kpa}.
The study of $t\bar{t}W$ production at very large invariant mass of
the $t\bar{t}$ system, furthermore, allows to probe directly the
$t\bar{t}g$ vertex in the domain of gluon virtuality $Q\sim
m_{tt}$. This is because the leading production diagram has a $W$
emitted from the initial state, followed by the $s$-channel
annihilation $q\bar{q}\to t\bar{t}$. 

The final state $t\bar{t}Z$, on
the other hand, allows to measure directly the $t\bar{t}Z$ vertex,
since this is the coupling that drives the dominant $gg\to t\bar{t}Z$
contribution\cite{Baur:2004uw}. 
Future LHC data will allow to set stringent and
model-independent limits on anomalous dipole contributions to the
$t\bar{t}Z$ vertex, with sensitivity comparable to that obtained from
indirect EW precision measurements at LEP~\cite{Rontsch:2015una}.

Both $t\bar{t}W$ and $t\bar{t}Z$ processes are known theoretically
with full NLO accuracy in
QCD\cite{Lazopoulos:2008de,Kardos:2011na,Campbell:2012dh,Garzelli:2012bn},
which leads to an intrinsic systematic uncertainty of about $\pm
10\%$.  In spite of low production rates at the energies of the first
run of the LHC ($\sigma(t\bar{t}W^\pm)\sim\sigma(t\bar{t}Z)\sim 200\pm
20$~fb at $\sqrt{S}=8~\tev$), ATLAS and CMS have nevertheless obtained
a signal evidence, for both processes, 
at the level of $5\sigma$, or better.  The first CMS
results\cite{Khachatryan:2014ewa} have been updated
recently\cite{Khachatryan:2015sha}, leading to the measurements of
$\sigma(t\bar{t}W^\pm)=382{{+117}\atop{-102}}$~fb (4.8$\sigma$) and
$\sigma(t\bar{t}Z)=242{{+65}\atop{-55}}$~fb (6.4$\sigma$).
ATLAS\cite{Aad:2015eua} measured
$\sigma(t\bar{t}W^\pm)=369{{+100}\atop{-91}}$~fb (5.0$\sigma$) and
$\sigma(t\bar{t}Z)=176{{+58}\atop{-52}}$~fb (4.2$\sigma$).  All these
results are well compatible with the SM predictions.

\section{Conclusions}
The production of vector gauge bosons in hadron collisions is like a
swiss knife: it is a versatile, reliable and robust tool for physics
at the high-energy frontier. It exposes a vast variety of phenomena,
covering most aspects of the dynamics of both EW and strong
interactions. While contributing to our deeper understanding and
consolidation of the SM, the knowledge acquired about the production
mechanisms of gauge bosons is  also essential to study the
properties of the top quark and of the Higgs boson, and to refine the
sensitivity of searches for BSM phenomena.

The precision of measurements and theoretical calculations has greatly
improved in the past few years, and allows now comparisons of gauge boson
production properties at the few percent level of precision. This precision is
superior to what can be achieved in most other hard processes in
hadronic collisions, and is liable to improve even further, through
continued theoretical and experimental efforts and additional
ingenuity. Already today, this precision can be used to improve the
determination of the EW parameters and of the proton PDFs, competing
with the pre-LHC state-of-the-art provided, respectively, by the
results of the LEP and SLC $e^+e^-$ colliders, and of the HERA $ep$
collider. It is easy to predict that the physics of $W$ and $Z$ bosons
at the LHC will continue surprising us for its richness for a long
time to come.

\printindex                         
\end{document}